\documentclass[aps,prl,twocolumn,superscriptaddress,floatfix]{revtex4-2}

\usepackage{graphicx,color}
\usepackage{amssymb}
\usepackage{amsmath}
\usepackage[export]{adjustbox}
\usepackage[breaklinks=true,colorlinks=true,linkcolor=blue,citecolor=blue,urlcolor=blue]{hyperref}
\usepackage{bbold,dsfont}

\def\Id{\mathbb{1}}
\newcommand{\beq}{\begin{equation}}
\newcommand{\eeq}{\end{equation}}
\newcommand{\ket}[1]{\mbox{$ | #1 \rangle $}}
\newcommand{\bra}[1]{\mbox{$ \langle #1 | $}}
\newcommand{\tr}{{\mathrm{tr}}}
\newcommand{\Hquad}{\hspace{0.5em}}
\newcommand{\alphaVec}{\boldsymbol{\alpha}}
\newcommand{\cVec}{\boldsymbol{c}}
\newcommand{\rhoVec}{\boldsymbol{\rho}}
\newcommand{\lambdaVec}{\boldsymbol{\lambda}}

\begin{document}

\title{Converting long-range entanglement into mixture: tensor-network approach to local equilibration}

\begin{abstract}

In the out-of-equilibrium evolution induced by a  quench, fast degrees of freedom generate  long-range entanglement that is hard to encode with standard tensor networks. However, local observables 
only sense such long-range correlations through their contribution to the reduced local state as a mixture. We present a tensor network method that identifies such long-range entanglement and efficiently transforms it into mixture, much easier to represent. In this way, we obtain an effective description of the time-evolved state as a
density matrix that captures the long-time behavior of local operators with finite computational resources.
\end{abstract}

\author{Miguel Fr{\'{\i}}as-P{\'e}rez}\email{miguel.frias@mpq.mpg.de}
\affiliation{Max-Planck-Institut f\"ur Quantenoptik, Hans-Kopfermann-Str.\ 1, D-85748 Garching, Germany}
\affiliation{Munich Center for Quantum Science and Technology (MCQST), Schellingstr. 4, D-80799 M\"unchen}
\author{Luca Tagliacozzo}
\affiliation{Institute of Fundamental Physics IFF-CSIC, Calle Serrano 113b, 28006 Madrid, Spain}
\author{Mari Carmen Ba\~nuls}
\affiliation{Max-Planck-Institut f\"ur Quantenoptik, Hans-Kopfermann-Str.\ 1, D-85748 Garching, Germany}
\affiliation{Munich Center for Quantum Science and Technology (MCQST), Schellingstr. 4, D-80799 M\"unchen}
\maketitle

\begin{figure}[t]
	\centering
	\includegraphics[width=\columnwidth]{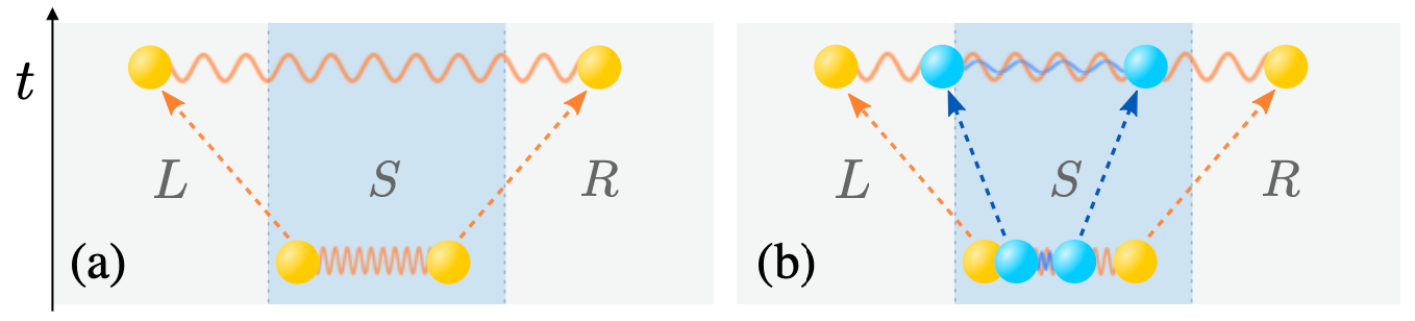}
 	\caption{(a) A correlated QP pair initially located inside region $S$ creates long-distance (pure $LR$) entanglement when both QPs propagate outside of $S$. (b) At that time, slower QPs (blue) can still contribute to entanglement between $S$ and its surroundings.}
	\label{fig:figQP}
\end{figure}

Reconciling the time-reversal invariant unitary evolution of closed quantum many-body systems with the emergence of statistical mechanics and its well defined arrow of time is still an open question, hindered by the exponential complexity of simulating this problem in the generic case.

In contrast, the equilibrium wave functions of a large class of physically relevant systems live in a small corner of the exponentially large Hilbert space, characterized by a bounded amount of entanglement~\cite{Eisert2010area}. States therein admit an efficient approximation as a tensor network (TN)~\cite{Cirac2021rmp}. This property implies we can perform very precise numerical simulations with only polynomial resources ~\cite{Verstraete2008,Schollwoeck2011,Orus2014annphys,Silvi2019tn,Ran2020tncontr,Okunishi2021review,Banuls2023ar}, such as those with matrix product states (MPS) in one dimension~\cite{Fannes1992,White1992,Oestlund1995,PerezGarcia2007,Schollwoeck2011}.

Out of equilibrium, instead, initially localized correlations can propagate over arbitrarily large distances.
As a result, the entanglement in the system increases rapidly~\cite{Calabrese2005,Osborne2006,Schuch2008entropy,lauchli2008,rossini2010,Acoleyen2013entanglementRates} and simple TN ansatzes like MPS have limited applicability. By focusing on the local description of the state, nevertheless, one can circumvent the exponential complexity originated from such non-local correlations. Significant simplifications using this approach have already been observed~\cite{Banuls2009, Leviatan2017, White2018, Hallam2019lyapunov, Krumnow2019, Pastori2019, Surace2019, Rams2020,  Rakovszky2020, Kvorning2021, Frias2022}.

Here we develop this idea and propose a new algorithm that explicitly identifies long-range entanglement in the system and trades it for mixture (see also~\cite{Surace2019}). In particular, our method focuses on separating fast and slow propagating degrees of freedom.
Already from an early time we 
observe that the fast degrees of freedom mediate non-local correlations whose effect on local observables is 
that of a statistical mixture. Using that knowledge, we devise a TN algorithm that provides an effective description of the time-evolved state as a density matrix, represented by a matrix product operator (MPO) with bounded bond dimension, and 
accurately captures the long-time behavior of local operators.

To benchmark the algorithm, we simulate quenches in the transverse field Ising model and its non-integrable generalization. We find that our results agree with those dictated by, respectively, the analytical solutions and the prediction of the diagonal ensemble~\cite{srednicki1994,gogolin2016,vidmar2016}.

\begin{figure*}[t]
	\centering
	\includegraphics[width=\linewidth]{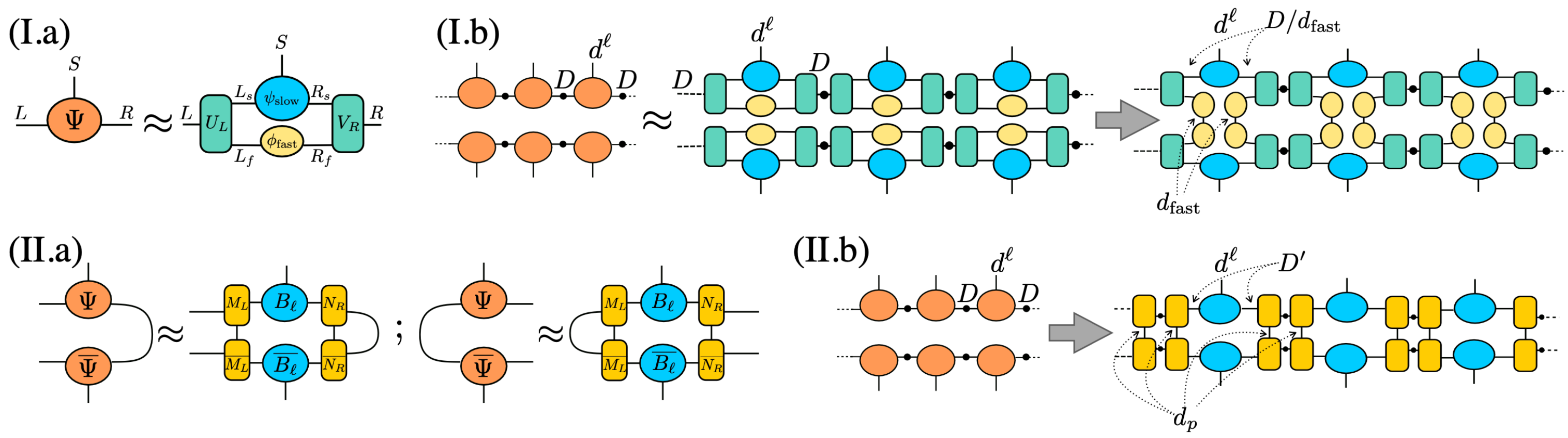}
	\caption{First iteration in the transformation of long-range entanglement into mixture in TN.
 (I.a) Decomposition~\eqref{eq:decomp} for a block $S$, represented by a single tensor; (I.b) The density matrix of the full state (left) 
 can be written in terms of the new tensors (middle) when we apply the decomposition to each block [black circles represent the inverse Schmidt values matrix, inserted to compensate the central gauge in (I.a)]. We then substitute each long-range component
 by the product of its marginals, giving rise to a mixed description (right).
 (II.a) The heuristic algorithm directly searches for tensors that (approximately) preserve the reduced density matrices for $LS$ and $SR$. (II.b) The density matrix for the full state is replaced by a purification defined by the solution.
 The structure is analogous to that in (I.b) (in each case we have indicated the relevant dimensions). 
}
	\label{fig:figTN}
\end{figure*}

\begin{figure}[t]
	\centering
	\includegraphics[width=.95\columnwidth]{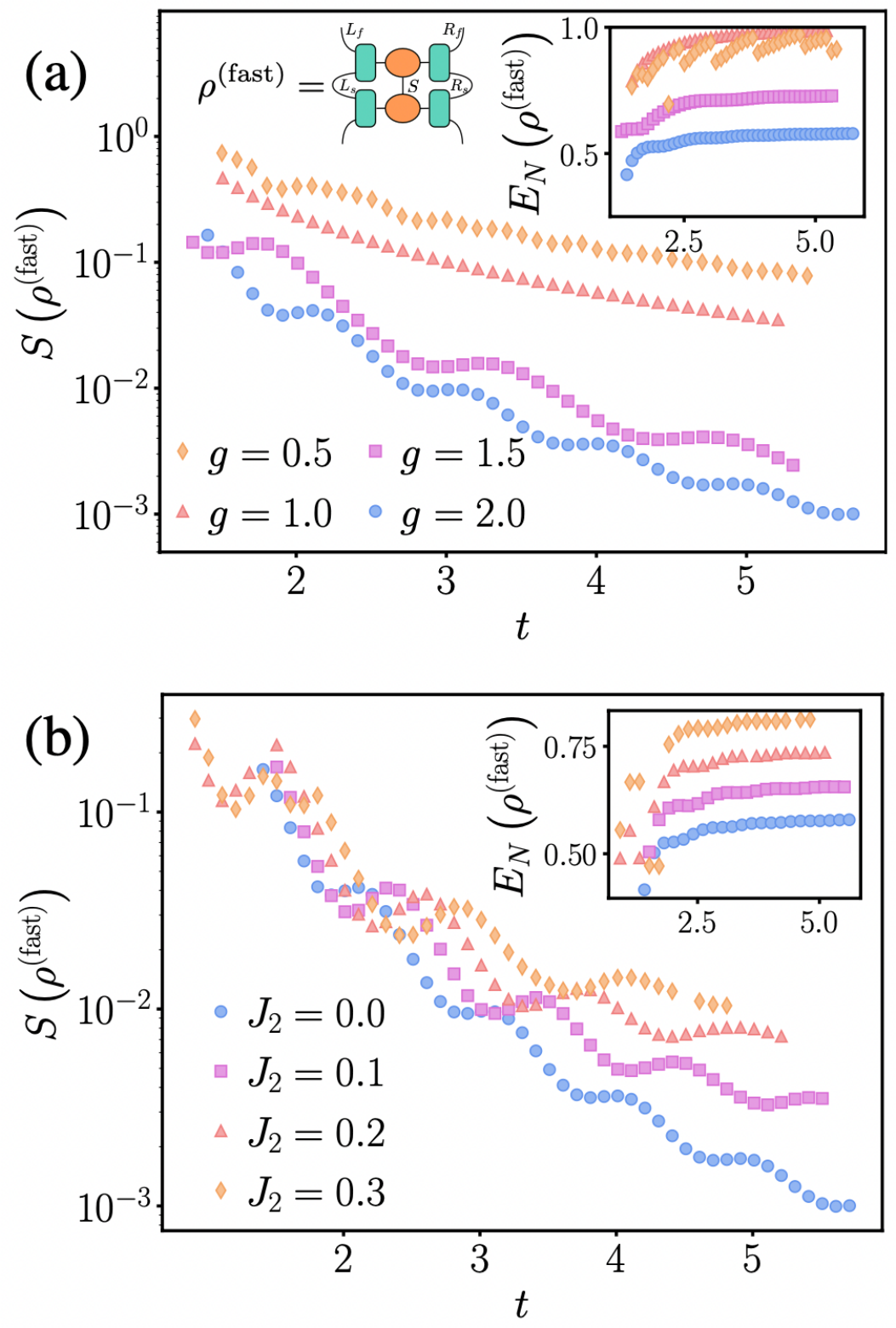}
	\caption{Entropy (main plots) and left-right logarithmic negativity (insets) of the reduced density matrix for the \emph{fast} degrees of freedom [see sketch in panel (a)], computed from the decomposition in fig.~\ref{fig:figTN}(I.a) at different times (a) for several integrable quenches ($g\neq0$, $J_2 = 0$) and (b) non-integrable ones ($g = 2$, $J_2\neq0$).}
\label{fig:figEntropy}
\end{figure}

\paragraph{Quenches and quasiparticles.}
We focus on quantum quenches ~\cite{Calabrese2005,Mitra2018quench}, in which a
one-dimensional system in the thermodynamic limit is prepared at $t=0$ in a product state $\ket{\psi_0}$, and
later evolved with an entangling Hamiltonian $H$. 
Our aim is to compute the time-dependent expectation value of a local observable $O$.

The initial state has a finite energy density above the ground state, and thus a large occupation of excited states, often described as quasiparticles  states (QP). The subsequent dynamics can be described as the radiation of entangled pairs of QPs ~\cite{Calabrese2005}.  In translational invariant systems indeed QPs possess a well-defined 
dispersion relation $\epsilon(k)$, and
a QP wave-packet centered at $k_0$ propagates with 
group velocity $v_{k_0}=\left.\partial_k \epsilon(k)\right|_{k_0}$, while momentum conservation enforces equal occupation of 
states that propagate with opposite momenta.

\paragraph{Identifying long-distance entanglement.}
In order to identify the long-distance contributions to the entanglement we need to focus on a subsystem $S$ of $\ell$ neighboring spins, and label $L$ and $R$ the remaining left and right regions of the system (see Fig.~\ref{fig:figQP}).
Entanglement across a bipartition (e.g. $L$ vs $SR$) is created by correlated pairs of QPs with support on both sides of the cut.
We define as \emph{long-distance entanglement} the contribution to the entanglement generated by  any QP pair supported on $L$ and $R$, but not on $S$.

We can visualize this in the cartoon of Fig.~\ref{fig:figQP}(a), where we consider a quench that excites a single pair of entangled QPs, initially located inside $S$,
that later travel with opposite velocities $\pm v$.\footnote{This kind of simple dynamics is actually realized in a toy model of local unitaries~\cite{Banuls2009,MullerHermes2012} and in some concrete physical realizations such as the split quenches studied in \cite{zamora2014splitting,chen2014}.}
In this simple example,
$S$ becomes entangled with the rest when one member of the pair leaves the region, but after $t\sim  \ell/v$ both QPs have left, and the total state factorizes into the product of an entangled state of $LR$ and a state of $S$.

In a more general scenario such complete factorization is not possible, since $S$ and $LR$ might contain other sources of entanglement.
The simplest generalization depicted in  Fig.~\ref{fig:figQP}(b) involves two pairs of QPs, one \emph{fast} (orange) and one \emph{slow} (blue). When the former has left $S$ the latter is still partially in $S$.
At that stage, it is still possible to disentangle a part of $L$ and $R$ from $S$.
We can indeed  identify a factorization of the Hilbert spaces, $L\equiv L_s \otimes L_f$ and $R \equiv R_s \otimes R_f$,
such that the subsystem $L_f\otimes R_f$ is not entangled with $S$. In the simple scenario presented above, $L_f\otimes R_f$ is  defined by the degrees of freedom that describe the fast pair, and we can factorize the state as $\ket{\psi_{L_sSR_s}^{\mathrm{(slow)}}}\otimes\ket{\phi_{L_fR_f}^{\mathrm{(fast)}}}$, where the second factor captures the pair of fast modes.

\paragraph{Trading long-distance entanglement for mixture.}
In a translationally invariant system we can repeat the cartoon picture above for each block of $\ell$ sites. When computing the expectation value of a local observable, 
all the degrees of freedom are traced out, except those supporting the observable. For fast entangled pairs separated by at least $\ell$ sites, one partner will necessarily be traced out, leaving the other in a mixed state.

Thus, when focusing on local observables,  we can describe the system as a mixed state where, for each fast pair, we have substituted the entangled state $\rho_{L_f R_f}^{\mathrm{(fast)}}$ by $\tilde{\rho}_{L_f R_f}^{\mathrm{(fast)}}$, the product of the mixed states obtained by tracing out each partner in the pair,
\begin{align}
\rho_{L_f R_f}^{\mathrm{(fast)}}=
&\ket{\phi_{L_f R_f}^{\mathrm{(fast)}}}\bra{\phi_{L_f R_f}^{\mathrm{(fast)}}} \rightarrow 
\tilde{\rho}_{L_f R_f}^{\mathrm{(fast)}} = \rho_{L_f}^{\mathrm{(fast)}} \otimes \rho_{R_f}^{\mathrm{(fast)}},
\label{eq:mix}
\end{align}
where $\rho_{L(R)}^{\mathrm{(fast)}} \equiv \tr_{R(L)} \left ( \rho_{LR}^{\mathrm{(fast)}} \right )$ are the reduced density matrices of each QP.
This provides a more efficient local description 
in terms of entanglement, as the long-distance components have been removed.

\paragraph{From the QP intuition to a TN algorithm. } 
Standard MPS algorithms for this setup 
attempt to represent the time-evolved state 
as a MPS, parametrized by one or few tensors $A_m$ of 
size $d\times D \times D$, where $d$ is 
the physical dimension of the chain sites and $D$ the bond dimension~\cite{Vidal2007,Haegeman2011tdvp,Vanderstraeten2019ln}.  Since the half-chain entropy of the MPS is upper-bounded by $\log D$~\cite{perez-garcia2007,Schollwoeck2011,evenbly2011},
the typical linear growth in time of entropy after a quench~\cite{Calabrese2005,Schuch2008entropy} requires an exponential increase of the tensors dimensions to maintains a constant precision.
As a result, given finite computational resources, standard TN algorithms ~\cite{Paeckel2019tevol}  only give reliable predictions for relatively short times (typically of the order $J t \simeq 10$ with $J$ the relevant energy scale).

Translating the QP intuition above to the TN setting we can however
obtain a more efficient TN description for the local observables.
Let us, for simplicity, assume that each MPS tensor represents precisely $\ell$ sites.\footnote{This can be always obtained by blocking together $\ell$ tensors of the original systems to yield a single tensor.}
Singling out one subsystem, the whole state can thus be written as
\beq
\ket{\Psi}=\sum_{\alpha s_{\ell} \beta}  C_{\alpha \beta}^{s_{\ell}} 
\ket{\Phi^{{L}}_{\alpha} }
\ket{s_{\ell}}
\ket{\Phi^{{R}}_{\beta}},
\label{eq:stateLSR}
\eeq
where the sum is over orthonormal bases of the block and its left and right environments $\{\ket{\Phi^{{L/R}}}\}$, meaning that we use the gauge in which  the tensor $C_{\alpha \beta}^{s_{\ell}}$ for the subsystem $\{\ket{s}\}$ is the orthogonality center of the MPS ~\cite{Vidal2007,PerezGarcia2007,Vanderstraeten2019,evenbly2022}.

If there are long-range entangled degrees of freedom that decouple from $S$, the state will have a product structure, with 
one component completely disentangled from the physical degrees of freedom. 
There will thus exist basis transformations (disentanglers~\cite{vidal2007entanglement,evenbly2009algorithms}) on $L$ and $R$ 
identifying the decomposition $L=L_f\otimes L_s$ and $R=R_f\otimes R_s$, such that
\beq
U_{L}^{\dagger}\otimes \Id_S \otimes V_{R}^{\dagger} \ket{\Psi} = \ket{\psi_{L_s S R_s}}\otimes \ket{\phi_{L_f R_f}},
\label{eq:decomp}
\eeq
as depicted in Fig.~\ref{fig:figTN}(I.a). If $U_{L}$ and $V_R$ exist, they can be determined minimizing the Euclidean distance between the left and right-hand side of~\eqref{eq:decomp}~\cite{Kraft2018}. Namely, given the evolved state in the form~\eqref{eq:stateLSR}, and for fixed dimension $d_{\mathrm{fast}}$ of the long-range component, we iteratively optimize each of the disentanglers $U_L$ and $V_R$ and the vectors $\ket{\psi_{L_s S R_s}}$ and $\ket{\phi_{L_f R_f}}$ until the distance converges. Since the dimension $d_{\mathrm{fast}}$ of the fast factors $L_f$ and $R_f$ is unknown, the procedure needs to be repeated for different trial values.

Assuming the above procedure succeeds, we now can transform  the identified long-range entanglement into mixture by applying the substitution~\eqref{eq:mix} to each consecutive subsystem in the chain.
As schematically shown in Fig.~\ref{fig:figTN}(I.b), this
exchanges the pure MPS description of the system 
by a mixed (purified) MPO with smaller bond dimension $D/d_{\mathrm{fast}}$, at the expense of additional (purification) indices.

 When the decomposition~\eqref{eq:decomp} is exact, this mixed state has the property that the reduced density matrices for $S_{\ell}R$ (resp. $LS_{\ell}$) are unchanged~\cite{SM}.
 As a consequence, also the reduced density matrix for two adjacent blocks, and thus all expectation values and correlation functions with support on up to $\ell$ sites are preserved. Larger supports can be achieved working with a larger $\ell$, at the cost of longer disentangling times.
 We can then continue the evolution with the original Hamiltonian acting on the physical indices. 
 Fast degrees of freedom will continue propagating entanglement through the subsystem $S$.
Thus, the step of finding and mixing the long-range entangled components is repeated periodically.

\begin{figure}[t]
	\centering
	\includegraphics[width=.85\columnwidth]{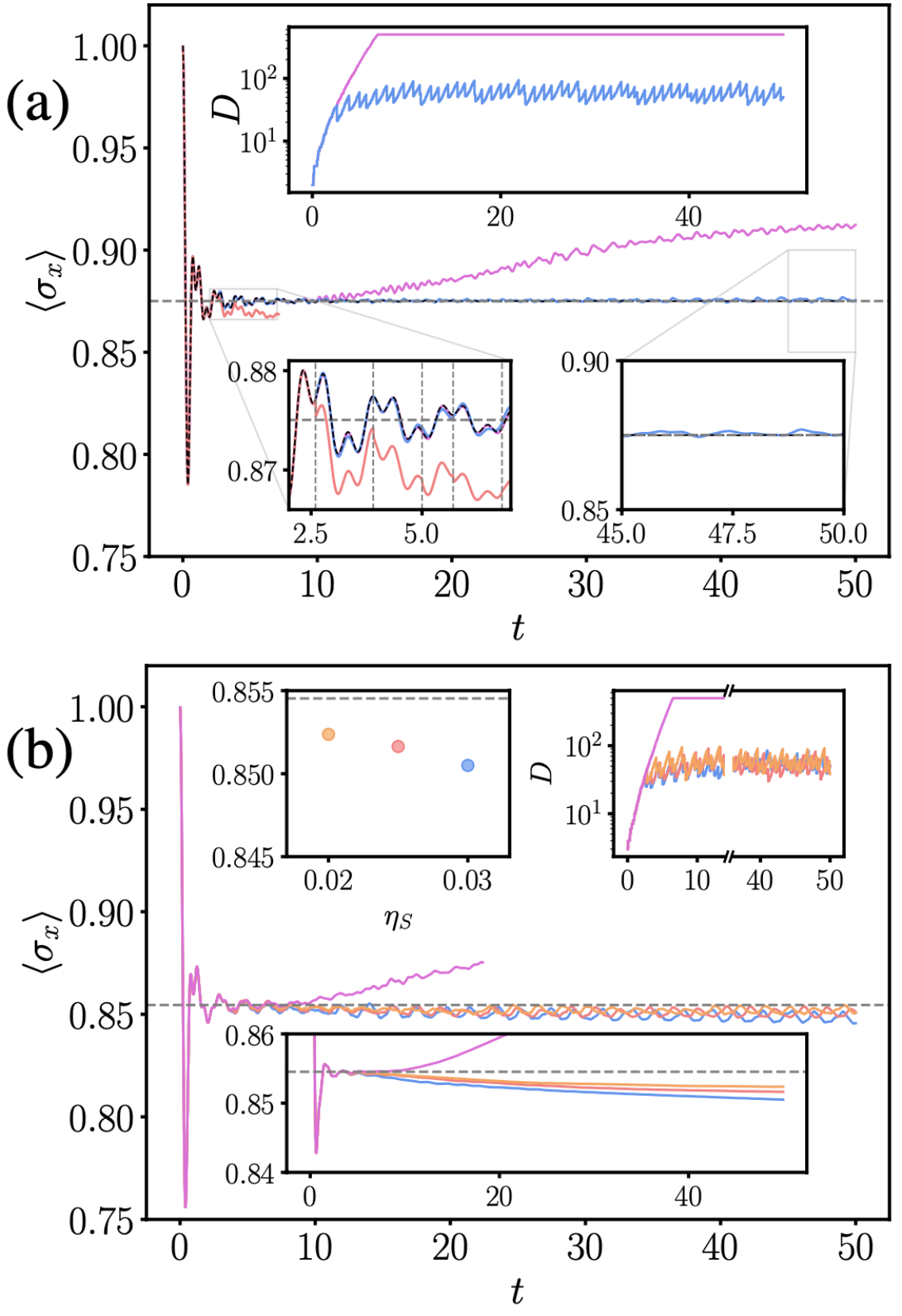}
	\caption{  (a) Evolution of the transverse magnetization for the integrable quench ($g=2$, $J_2=0$) as obtained with the direct decomposition of fig.~\ref{fig:figTN}(I) (orange) and the heuristic algorithm of fig.~\ref{fig:figTN}(II) (blue). For comparison, we show the exact result (dashed black line), the equilibration value (dashed grey) and the iTEBD result with $D=500$ (pink).  The lower insets show close-ups of the short- and long-time regions. In the left one, the vertical dashed lines indicate mixing steps, inducing jumps in the simple but not the heuristic algorithm. (b) Same observable for the non-integrable quench ($g=2$, $J_2=0.1$) using the heuristic algorithm. Different colors indicate different residual entropy thresholds $\eta_S$. Our simulations converge towards the diagonal ensemble value $\langle \sigma_x\rangle_{\mathrm{DE}}$ (dashed gray line). This can be more clearly seen in the time average of the observable  (lower inset) or the value of the time average at time $t=50$ as a function of $\eta_S$ (upper-left inset).
 For both quenches, the upper (right) inset shows the bond dimension as a function of time. }
 \label{fig:figNum}
 \end{figure}

\paragraph{An improved TN algorithm. }
In practice, the best decomposition we identify using this strategy retains some residual entropy between fast and slow degrees of freedom $S(\rho_{\mathrm{fast}})$. We apply the truncation when this entropy falls below a given threshold $\eta_S$.
This results into small errors when building the mixed state that reflect in the evolution of local observables [see Fig.~\ref{fig:figNum}(a)].
Therefore we introduce an improved heuristic TN algorithm in which, 
after solving~\eqref{eq:decomp}, if 
$S(\rho_{\mathrm{fast}})<\eta_S$ we propose an effective purification ansatz, described by three rank-3 tensors $M_L,\,B_{\ell},\,N_R$ with fixed dimensions $d_p DD'$, $d^{\ell} D'^2$ and $d_p D' D$, where $d_p$ is a chosen dimension for the purification index and $D'=D/d_p$ [see Fig.~\ref{fig:figTN}(II)].

The purification tensors are found variationally by minimizing the distance between the reduced density matrices $\rho_{SR}$ (and $\rho_{LS}$) obtained from the original state~\eqref{eq:stateLSR} and from the purification. As initial guess we use the tensors obtained from solving~\eqref{eq:decomp}, and we use gradient descent for the minimization problem (see~\cite{SM}).

The dimension of the purification indices doubles after each mixing step. In practice, after few steps we keep it fixed by discarding the smallest eigenvalues of its reduced density matrix since we observe that the error induced by this truncation is negligible compared to other sources of errors in the simulation.\footnote{Other schemes would be possible to try and optimize the purification sites, for instance, in the spirit of the optimal purification of~\cite{Hauschild2018puri}.}

\paragraph{Numerical results.}

We benchmark the algorithm using 
the transverse field Ising model with an additional integrability-breaking next-to-nearest-neighbor interaction, 
\beq
H= - \sum_i \left( \sigma_i^z \sigma_{i+1}^z + g \sigma_i^x + J_2 \sigma_i^z \sigma_{i+2}^z \right).
\label{eq:model}
\eeq
Initially in a product state $\ket{X+}\equiv \otimes\frac{1}{\sqrt{2}}\left(\ket{0}+\ket{1}\right)$, corresponding to the $g \to \infty$ ground state, 
the system is quenched to finite values of $g$, and potentially $J_2$.

To explore systematically the structure of the time-evolved MPS, we simulate accurately the short-time evolution
using the infinite time-evolving block decimation (iTEBD) algorithm~\cite{Vidal2007} with a large enough bond dimension.
We then probe the disentangling of fast and slow degrees of freedom at different times solving~\eqref{eq:decomp} 
for a subsystem of $\ell=2$ sites for several quenches.

As shown in Fig.~\ref{fig:figEntropy}, 
after a certain time 
we can identify
subsystems of $L$ and $R$ 
that practically disentangle from the local region and carry long-range entanglement between the environments, measured
by the logarithmic negativity~\cite{Vidal2002, Plenio} of the reduced density matrix for the fast degrees of freedom
$\tr_{L_s S R_s} \ket{\Psi} \bra{\Psi}$.
For all quenches, the residual entropy $S(\rho_{\mathrm{fast}})$ decays fast with time, with slower rate when quenching into the ordered phase.\footnote{This is consistent with the distribution of momentum excitations in the initial state, computed in the exact free-fermionic formulation, which is peaked at finite velocities when
the quench is within the same phase, and at zero velocity otherwise (see the Supplemental Material~\cite{SM}).} As expected, the time at which fast degrees of freedom start decoupling from the local system depends linearly on $\ell$ [see Fig.~\ref{fig:figBlockSize} in the Supplemental Material~\cite{SM}].

We now use the TN algorithm described above and check its performance simulating the long-time evolution of local observables.
Fig.~\ref{fig:figNum}(a) shows the results for the integrable quench $g=2$, $J_2 =0$
and the only non-vanishing (due to the $Z_2$ symmetry) single-site observable $\langle \sigma_x \rangle $.
 Solving \eqref{eq:decomp} and trading entanglement by mixture captures the qualitatively correct dynamics, but introduces discontinuous jumps 
(orange line and lower-left inset) that we attribute to the decomposition not being exact.
The improved heuristic  algorithm achieves much better results (blue line and lower-right inset).
While the approximation induces some residual oscillations, these are very small and the results are  close to the exact equilibration even after considerably long times.
In contrast, iTEBD results with a maximal bond dimension $D=500$ (pink line) start to severely deviate from the exact solution
at around $t=10$. 

Our algorithm uses instead a much smaller bond dimension: during the time evolution steps, the required bond dimension of the purified MPS  grows exponentially, but each time we trade some of the entanglement for mixture, the bond dimension gets halved.\footnote{We use $d_p=2$ for our ansatz.}
As we iterate the procedure, the maximum required bond dimension tends to a constant $D\sim100$ (upper inset).

In Fig.~\ref{fig:figNum}(b) we repeat a similar analysis for the non-integrable quench  $g=2$, $J_2=0.1$.
The different colors show the results for different thresholds for the residual entropy $\eta_S$. 
Smaller values require longer evolution between mixing steps, and thus larger bond dimension,
but improve the results systematically. 

Our results systematically approach the long-time limit predicted by the diagonal ensemble $\langle \sigma_x\rangle_{\mathrm{DE}}=0.852$.\footnote{Extrapolated from exact finite size results.}

Even for the largest threshold,
the relative error 
in the time-averaged value 
 after $t\sim 50$ is below $1\%$ ($0.848$). Also the time-averaged magnetization exhibits a similar precision, as shown in the lower inset.
Also in this case the largest bond dimension saturates with $t$ (upper right inset). 

In the Supplemental Material~\cite{SM}, we also repeat the calculations in the integrable case directly using the free-fermionic formalism~\cite{Surace2022} where we can track the coherence we discard in our simulation and confirm it does not play a relevant role in the long-time dynamics of local observables.

\paragraph{Discussion.}

By identifying the long-range entanglement produced in the out-of-equilibrium dynamics after a quantum quench and converting it into mixture, we have proposed an explicit approach to avoid the entanglement barrier and simulate the out-of-equilibrium dynamics of an infinite quantum chain with MPS using finite computational resources.

Our approach is inspired by the intuitive understanding of entanglement dynamics in terms of the radiation of QP. It relies on the hypothesis that fast degrees of freedom propagate correlations to steadily growing distances, contributing to the linear growth of entanglement across the system, but only as statistical mixture to sufficiently local observables. Our numerical results for the Ising chain show that the intuition is accurate in the free-fermionic case, which best fits the QP picture.

We have generalized our intuition to a heuristic algorithm that goes beyond the QP picture and also performs
well in non-integrable regimes of the model.
It is thus important to pursue further characterization of the algorithm in order to chart its potential and limitations.

\begin{acknowledgments}
LT would like to thank Jacopo Surace who has contributed to developing the original ideas that have motivated this work. 
This work was partially supported by 
the Deutsche Forschungsgemeinschaft (DFG, German Research Foundation) under Germany's Excellence Strategy -- EXC-2111 -- 390814868; 499180199  
and by the EU-QUANTERA project TNiSQ (BA 6059/1-1).
LT acknowledges support from the Proyecto Sin\'ergico CAM 2020 Y2020/TCS-6545 (NanoQuCo-CM), the CSIC Research Platform on Quantum Technologies PTI-001 and from Spanish projects PID2021-127968NB-I00 and TED2021-130552B-C22 funded by MCIN/AEI/10.13039/501100011033/FEDER, UE and MCIN/AEI/10.13039/501100011033, respectively.

\end{acknowledgments}

\bibliographystyle{apsrev4-2}
\bibliography{entanglementAndDynamics}

\clearpage
\onecolumngrid
\begin{center}
\textbf{\large Supplemental Material for\\[0.5ex]
``Converting long-range entanglement into mixture: tensor-network approach to local equilibration''}\\[2ex]
Miguel Fr{\'{\i}}as-P{\'e}rez, Luca Tagliacozzo, and Mari Carmen Ba\~nuls
\end{center}
\vspace{2ex}
\twocolumngrid

\setcounter{section}{0}
\setcounter{equation}{0}
\setcounter{figure}{0}
\renewcommand{\thesection}{S\Roman{section}}
\renewcommand{\theequation}{S\arabic{equation}}
\renewcommand{\thefigure}{S\arabic{figure}}

\section{Description of the algorithms}

In the main text we discussed a way to transform the long-range entanglement generated in a quench into mixture using tensor networks. The transformation is such that it reduces the bond dimension of the TN, while aiming to preserve the local observables and their evolution. In this section of the supplementary material we provide a complete description of the algorithms we use for that purpose. First, in order to set the notation and introduce some useful concepts, we briefly describe the uniform MPS formalism.
For a more complete description, we refer the reader to the excellent review~\cite{Vanderstraeten2019}, whose notation we follow here.

\subsection{Uniform MPS}

Uniform MPS (uMPS) are TN states that represent one-dimensional quantum many-body states in the thermodynamic limit. These states are constructed by repeating infinitely many times a tensor along a line (or repeating in a cyclic manner a finite set of them, if we work with a unit cell larger than one). In graphical TN notation, 

\beq
\ket{\Psi(A)}
 = \includegraphics[scale = 0.6, valign = m]{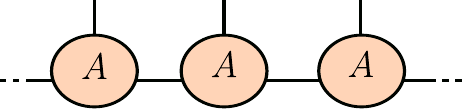}.
 \label{eq:uniformMPS}
\eeq
\noindent 
Where $A$ is a $d \times D \times D$ tensor, $d$ being the physical dimension of the sites of the chain and $D$ the bond dimension. A central object in uMPS simulations is the transfer operator, defined as the following four-legged tensor:
\beq
E(A)=
	\includegraphics[scale = 0.6, valign = m]{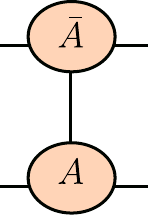} \quad .
 \label{eq:transferOperator}
\eeq
\noindent 
We can interpret this object as a $D^2 \times D^2$ matrix
(left vs. right indices). Many of the important properties of the state are captured by the dominant left and right eigenvectors of this matrix.
For instance, they determine the reduced density matrix of any (connected) subsystem and thus appear in the calculation of any local expectation value.
In the generic case, these dominant eigenvectors are non-degenerate~\cite{PerezGarcia2007}. If the state is normalized, the largest eigenvalue is 1, and the leading eigenvectors  correspond to the fixed points of the transfer operator seen as a map from the left to the right indices and viceversa.

It is possible to exploit the gauge freedom in the TN to transform these eigenvectors into a 
canonical form, which is particularly simple.
In particular, in the so-called \emph{left-canonical} form, the MPS tensor for a normalized state satisfies the conditions
\beq
    \includegraphics[scale = 0.6, valign = m]{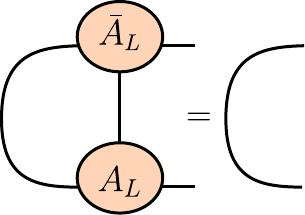}, \quad
        \includegraphics[scale = 0.6, valign = m]{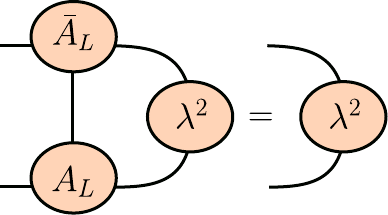} \Hquad,
	\label{eq:fig_canonical_form}
\eeq
where $\lambda^2$ is a diagonal matrix with positive entries. 
Cutting~\eqref{eq:uniformMPS} in half across one cut defines $D$ states for the left half-chain $\ket{\Phi_{\alpha}^L}$, and the same number for the right half-chain, indexed by the (now open) virtual leg. The first condition in~\eqref{eq:fig_canonical_form} ensures that the $D$ states on the left half-chain
are orthonormal.
Alternatively, one can impose a \emph{right-canonical} form $A_R$ in which the conditions are the symmetrical ones and orthonormality is satisfied by the states to the right of the cut.

In practice, it is instead common to work in a \emph{mixed} gauge, in which one can define~\cite{Vanderstraeten2019} a single central tensor $C^{s}_{\alpha \beta} = \sum_\gamma (A_L)^s_{\alpha \gamma} \lambda_{\gamma \beta}$, where $\alpha$ ($\gamma$) is the left (right) virtual index of the tensor $A_L$, and $s$ denotes its physical index. The state can then be written as a half-chain of left-canonical tensors to the left of $C$ and a half-chain of right-canonical ones to the right, such that both left and right virtual bonds of $C$ are in orthonormal bases. Taking the tensor product of these two bases with the single-site physical basis, the state can be written as,
\beq
\ket{\Psi}=\sum_{\alpha s \beta}  C_{\alpha \beta}^{s} 
\ket{\Phi^{{L}}_{\alpha} }
\ket{s}
\ket{\Phi^{{R}}_{\beta}}.
\label{eq:stateCentral}
\eeq
We can rewrite the full uMPS in terms of the single tensor C, as
\beq
	\includegraphics[scale = 0.6, valign = m]{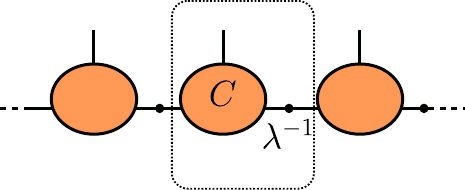} \quad .
 \label{eq:c_lambda}
\eeq

In this work, we are interested in the long-range entanglement between distant regions of the system $L$ and $R$, separated by a block $S$ of finite size $\ell$.
Thus, it is convenient to use the central canonical form with respect to the whole block $S$, such that 
we can study the long-range entanglement at the level of the virtual indices of the MPS.  
After blocking the $\ell$ sites of the subsystem $S$, the wave function of the system can be written as
\beq
\ket{\Psi}=\sum_{\alpha s_{\ell} \beta}  C_{\alpha \beta}^{s_{\ell}} 
\ket{\Phi^{{L}}_{\alpha} }
\ket{s_{\ell}}
\ket{\Phi^{{R}}_{\beta}}
\label{eq:stateLSR1}.
\eeq

As we will see in the next three subsections, our study of the long-range entanglement will allow us to construct a low-rank decomposition of the tensor $C$ when partitioned between $L$ and $SR$ (or $LS$ versus $R$), at the cost of introducing indices that couple the tensors in the bra and the ket. This decomposition will be such that the local expectation values and their evolution will be preserved, while the long-range coherences will be discarded.

\subsection{Detection of long range entanglement}

In this subsection we describe the algorithm that we use to identify the long-range entanglement in our MPS.

As discussed in the main text, if there are long-range entangled degrees of freedom that decouple from $S$, the state will 
factorize in a
tensor product 
with one component completely disentangled from the physical degrees of freedom in $S$. There will thus exist basis transformations on $L$ and $R$ identifying the decomposition $L = L_f \otimes L_s$ and $R = R_f \otimes R_s$, such that the degrees of freedom in the Hilbert spaces $L_f$ and $R_f$ are in a tensor product structure with the rest of the system. Graphically, 
\beq
\vcenter{\hbox{\includegraphics[scale = 0.6, valign = m]{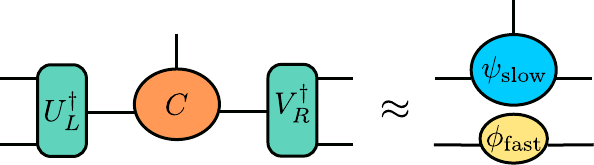}}}
\Hquad . 
\label{eq:simple_deco}
\eeq
\noindent
To look for such a decomposition we numerically minimize the square of the Euclidean distance between the left and the right-hand side of the previous diagram~\eqref{eq:simple_deco}. The variables in the optimization problem are the unitary matrices  $U_L$ and $V_R$, which respectively
determine the left and right basis transformations,
and the two resulting vectors, $\ket{\psi_{L_s S R_s}}$ and $\ket{\phi_{L_f R_f}}$. 
The full optimization is a complicated non-linear problem. However, as a function of each of the individual pieces, the cost function is quadratic and can be solved using basic linear algebra. Hence, to find the decomposition we use an alternating scheme, first used in~\cite{Kraft2018}: we fix all the variational tensors but one, and substitute the free piece by the one that minimizes the value of the cost function. When the unitaries are fixed, the optimal for the two vectors are just the leading left and right singular vectors across the partition $L_s S R_s$ versus $L_f R_f$. For the case of the unitaries, after imposing the unitarity constraint, the cost function is linear in each of them. That kind of problems can also be easily solved with a singular value decomposition of the environment~\cite{evenbly2009algorithms}.

The algorithm is efficient, with cost scaling  as $O(D^3)$
with the bond dimension, and we find that it shows a good performance.
In particular, we find components of the wave function which mediate long-range entanglement and decouple from the slower degrees of freedom, as expected according to the intuition drawn from the quasiparticle picture.

\subsection{Simple truncation algorithm}

In the previous section we have discussed the TN algorithm we use to identify the degrees of freedom that contribute to the long-range entanglement in our MPS representation of the time-evolved state. In practice, we 
combine
the previous algorithm 
with an iTEBD simulation that provides us with a faithful representation of the state at short times and later allows us to evolve the local tensors in the ansatz. 

\begin{figure}
	\centering
	\includegraphics[scale = 0.6]{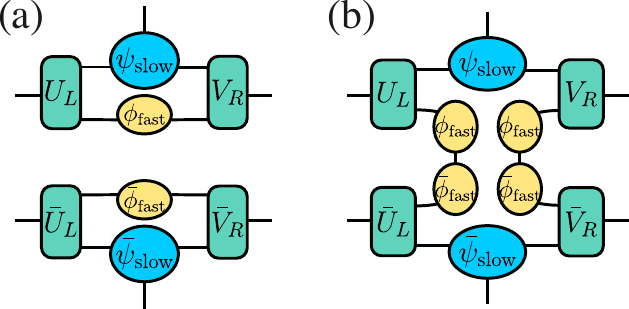}
     \caption{Graphical representation of the density matrices of the entire state before (a) and after the truncation of the fast degrees of freedom (b).}
	\label{fig:fig_simple_truncation}
\end{figure}

Starting from a pure state (uMPS) description of the initial state, we evolve unitarily with iTEBD. At periodic time intervals, we look for the above decomposition of the simulated wave function. The extent to which our time-evolved state has the sought structure can be characterized in several ways. We  choose to use the entanglement entropy between the fast and slow degrees of freedom, which can be identified after we have transformed the basis of the bond dimension indices according to the found unitaries.
We thus need to compute the entropy of the reduced density matrix $\rho_{L_f R_f}^{\mathrm{(fast)}} = \tr_{L_s S R_s} \ket{\Psi} \bra{\Psi}$, which can be graphically represented as
\beq
\rho^{\mathrm{(fast)}}=
 \vcenter{\hbox{ \includegraphics[scale = 0.6, valign = m]{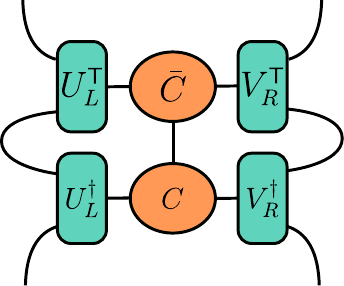}}} \Hquad . 
\eeq
\noindent
A vanishing entanglement entropy means that our state has, in the found virtual bases, exactly the desired tensor product structure, $\ket{\Psi}=\ket{\psi_{L_sSR_s}^{\mathrm{(slow)}}}\otimes\ket{\phi_{L_fR_f}^{\mathrm{(fast)}}}$. In this case, it is possible to modify the state in such a way that local expectation values are preserved while discarding some long-range correlations, as described in the main text. 
More concretely, we substitute the density operator of the full system $\ket{\Psi}\bra{\Psi}$, represented graphically in Fig.~\ref{fig:fig_simple_truncation} (left),
by a mixed state (right) that, in the basis of the virtual degrees of freedom in which the state factorizes, corresponds to 
$$
\ket{\psi_{L_sSR_s}^{\mathrm{(slow)}}}
\bra{\psi_{L_sSR_s}^{\mathrm{(slow)}}}
\otimes
\rho_{L_f}^{\mathrm{(fast)}} \otimes \rho_{R_f}^{\mathrm{(fast)}},$$
with $\rho_{L(R)}^{\mathrm{(fast)}} \equiv \tr_{R(L)} \left [ \rho_{LR}^{\mathrm{(fast)}} \right ]$ as defined in the main text.

\begin{figure}[h]
	\centering
	\includegraphics[scale=0.6]{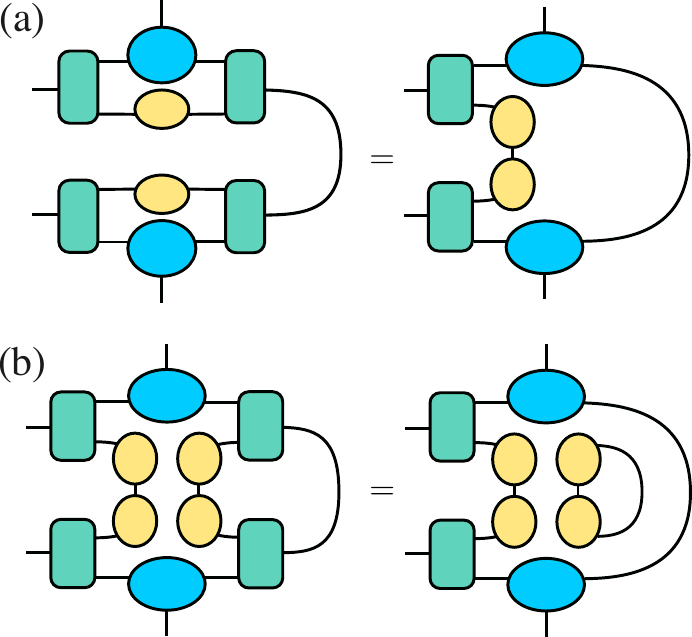}
    \caption{Reduced density matrix on the subsystem $LS$ before (a) and after (b) the truncation. Using the unitarity of $V_R$, one can check that the reduced density matrices are identical (right-hand side of the equalities). The state $\ket{\phi_{L_fR_f}^{\mathrm{(fast)}}}$ is normalized, and thus, the tensor contraction in the right-hand side of the equation in (b) gives 1. This same diagram also shows that the fixed points are unchanged by the mixing step: notice that by tracing subsystem $R$, we get a condition equivalent to the second equality in Eq.~\eqref{eq:fig_canonical_form} (namely, that $\sum_{s_\ell \beta} C^{s_\ell}_{\alpha \beta} \bar{C}^{s_\ell}_{\alpha' \beta} = (\lambda^2)_{\alpha \alpha'}$). Both the state before and after the truncation give the same resulting tensor. A similar argument can be made for the left fixed point.}
	\label{fig:fig_reduced_LS}
\end{figure}

The state on Fig.~\ref{fig:fig_simple_truncation}(a) is an exact representation of the tensor $C$ in the case in which the time-evolved state presents the exact sought structure. The state on the right (Fig.~\ref{fig:fig_simple_truncation}(b)) has a different structure: it represents a mixed state. Despite the two states being globally different, the two have the same reduced density matrices on the subsystems $L S$, as graphically shown in Fig.~\ref{fig:fig_reduced_LS}. An analogous computation shows that the reduced density matrix for $S R$ is also preserved by this substitution. 
Furthermore, as can be seen from  Fig.~\ref{fig:fig_simple_truncation}, the rank of the state (or bond dimension) across the cut $L$ versus $SR$ is reduced by a factor which is given by the dimension of the fast degrees of freedom.

\begin{figure}[h]
	\centering
	\includegraphics[scale=0.43]{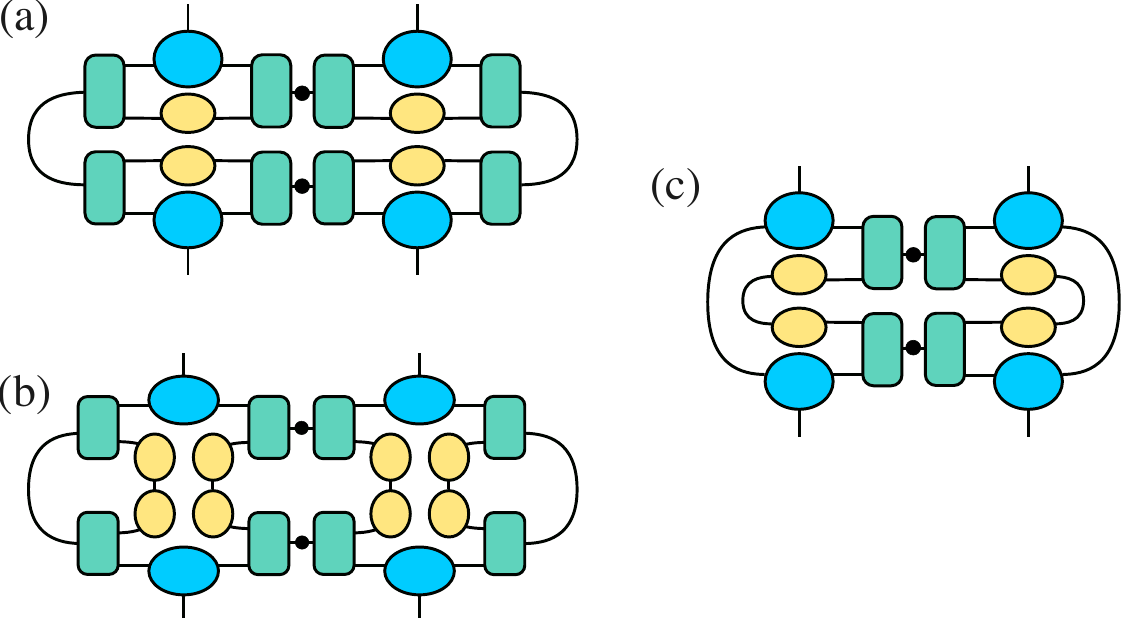}
    \caption{Reduced density matrices on two consecutive blocks of the system, before (a) and after (b) the truncation of the fast degrees of freedom. As explained in the text, the fixed points are also unchanged by the transformation. Using the unitarity of the two basis transformations in the virtual degrees of freedom, one can show that the expectation value for both is given by the same contraction (c).}
	\label{fig:obs}
\end{figure}

\begin{figure}[h]
	\centering
	\includegraphics[width=\columnwidth]{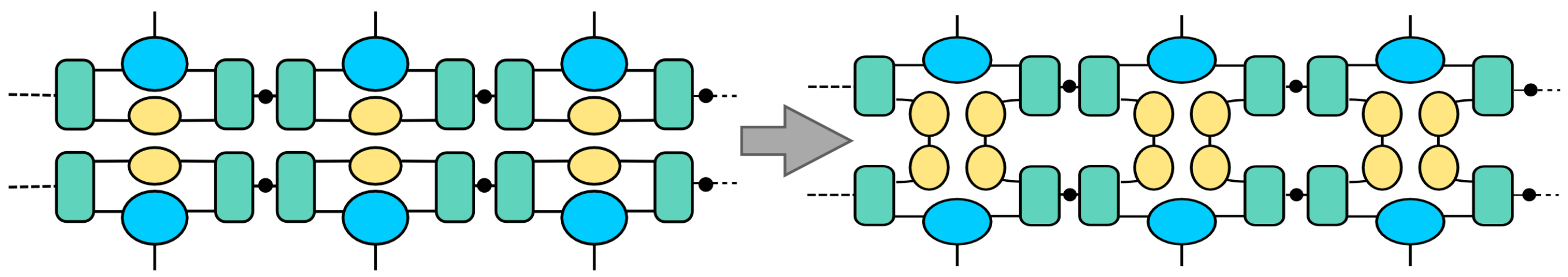}
    \caption{After identifying the long range entanglement, we truncate the fast degrees of freedom in all block simultaneously.}
    \label{fig:simple_truncation_all_sites}
\end{figure}

Because the state of the chain can also be explicitly written as a sequence of blocks, using the form~\eqref{eq:c_lambda}, we can perform the same transformation in all blocks of the chain at the same time [i.e., we substitute all the tensors $C$ in~\eqref{eq:c_lambda} as shown in Fig.~\ref{fig:simple_truncation_all_sites}]. After this substitution, it can be shown that the leading left and right eigenvectors of the transfer operator for the block are unchanged (see Fig.~\ref{fig:fig_reduced_LS}). This can be used to show graphically that the reduced density matrices of up to two blocks ($2\ell$ consecutive sites) are preserved (Fig.~\ref{fig:obs}).

\begin{figure}[t]
	\centering
	\includegraphics[width=.85\columnwidth]{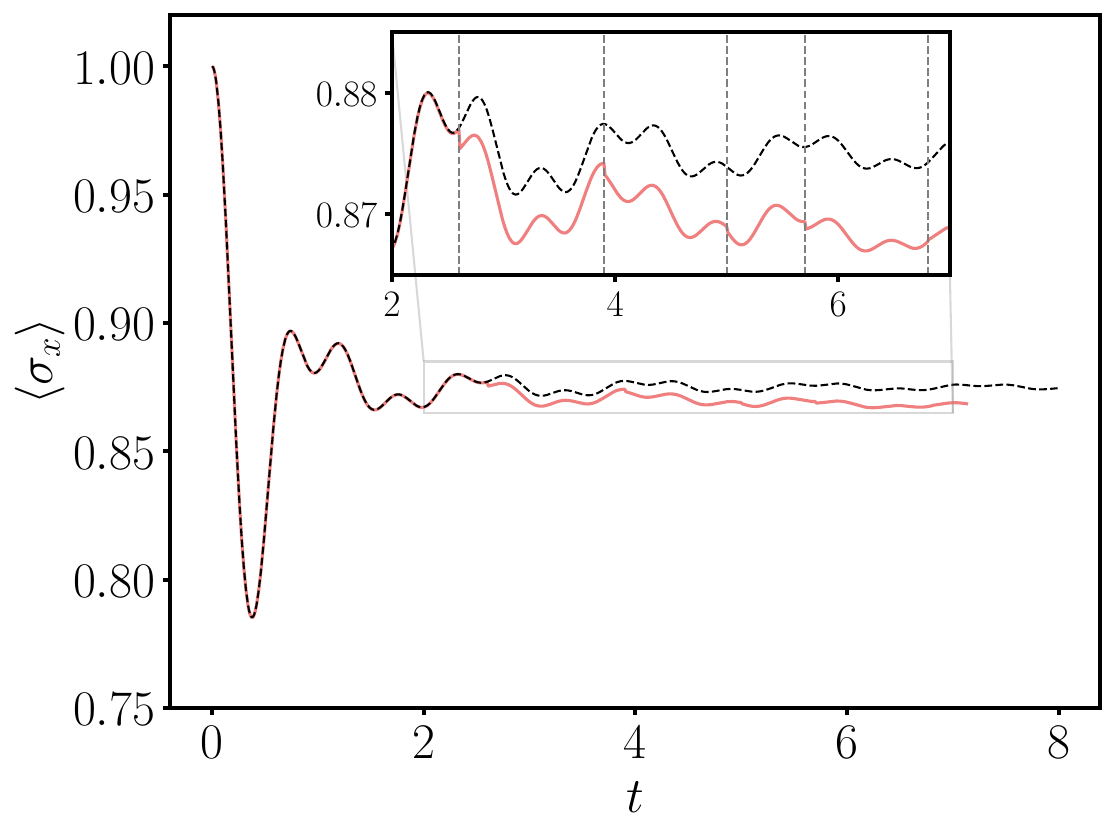}
	\caption{Evolution of the transverse magnetization for the integrable quench ($g = 2, J_2 = 0$) as obtained with the algorithm discussed in the section above, in orange. The exact result is shown by the black dashed line. The inset shows a close-up of the evolution. In the inset, the vertical gray dashed lines show the times in which the truncation is performed.
 }
 \label{fig:figSimple}
 \end{figure}

In practice, as can be seen from Fig.~\ref{fig:figEntropy} of the main text, the decomposition we obtain variationally never reaches full separation between fast and slow degrees of freedom and as a result the substitution is never exact. However, the fast-slow entropy seems to decay fast with time. Our simple truncation consists just on performing an iTEBD simulation, while monitoring the ``disentangling'' entanglement entropy. When it falls below a certain threshold, we substitute all the $C$ tensors on the present state by its tensor product approximation, and mix the fast degrees of freedom according to the diagrams above. After that, because the last step is not exact, we normalize again the state. The results from this evolution are the orange line in Fig.~\ref{fig:figNum}(a) of the main text. In them, it can be appreciated that at the times where the truncation is performed, we incur in an instantaneous error in the local observable. However, as we evolve the state again the dynamics seems to be correctly captured modulo the instantaneous shift introduced by the truncation. The same phenomenon repeats itself as we perform more and more truncations in the same simulation. A closer-up figure of the results for this truncation in the main text can be seen in Fig. \ref{fig:figSimple}.
The qualitative agreement with the true dynamics of this simple algorithm
corroborates the intuitive picture discussed in the main text, and 
justifies using these tensors as starting point for an improved method that corrects for those discontinuous jumps in a heuristic manner. We discuss the improved technique in the next section.

\subsection{Heuristic truncation}

Inspired by the results of our initial truncation, we present in this section a small modification of the previous algorithm that attempts to correct the shifts induced by the fact that we do not find an exact factorization between fast and slow degrees of freedom. 

\begin{figure}
	\centering
	\includegraphics[scale = 0.5]{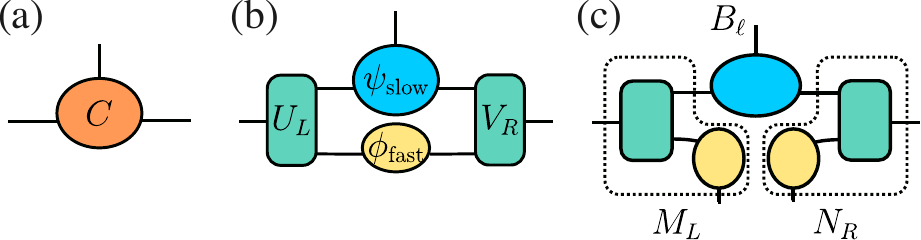}
	\caption{(a) Tensor $C$ obtained from the uMPS dynamical simulation in the mixed gauge. (b) Using the entanglement decomposition, we find an approximation to the tensor $C$ in terms of the different pieces that mediate the fast and slow degrees of freedom. (c) We group the pieces according to the dotted lines to construct the new tensors $M_L$, $B_\ell$ and $N_R$, which will be the starting point for the heuristic optimization variables.}
 \label{fig:schemeHeuristic}
 \end{figure}

\begin{figure}
	\centering
	\includegraphics[width=\columnwidth]{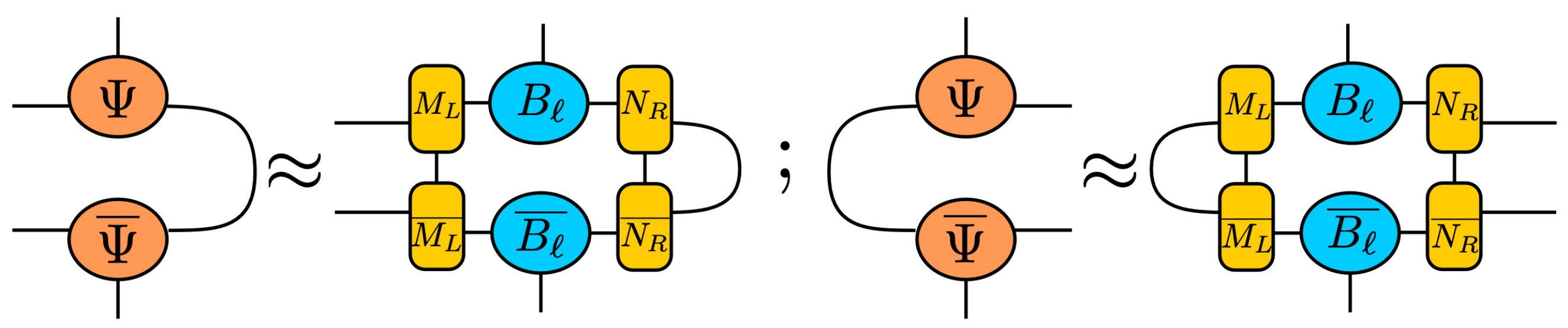}
	\caption{In order to improve our purification, we minimize the squares of the euclidean distances between the left and the right hand side of the two approximate equations above. We minimize the sum of the two distances, as we need a resulting state that satisfies both.}
 \label{fig:heuristicConditions}
 \end{figure}

Our improved heuristic algorithm 
takes advantage of this fact and tries to find an alternative purification form that generalizes the one of the simple algorithm and guarantees the correct marginals for $LS$ and $SR$.
The ansatz we use is composed by three tensors $M_L, B_\ell$ and $N_R$. We initialize their values from the pieces of the decomposition recovered by the simple algorithm and modify them to ensure the desired property (see Fig.~\ref{fig:schemeHeuristic} and Fig.~\ref{fig:heuristicConditions}). 

In order to improve the purification, we minimize the sum of the Euclidean distances between the new and previous reduced density matrices, namely we are interested in
$$\mathrm{argmin}_{M_L,B_{\ell},N_R} \left (\lVert \rho_{LS} - \tilde{\rho}_{LS}\rVert^2 + \lVert \rho_{SR} - \tilde{\rho}_{SR} \rVert^2 \right ),$$ 
where $\tilde{\rho}$ denotes the new reduced density matrices obtained from the mixed state defined by the tensors $M_L, B_\ell$ and $N_R$, which are the variables in the optimization. To minimize the cost function we use a gradient-descent scheme, as the optimization contains terms which are of fourth order in the entries of the tensors. The dimensions of the tensors, and the fact that they have purification legs guarantee that the original pure state will be approximated by a mixed state with a lower bond dimension.

The same as in the simple algorithm, we combine this decomposition with an iTEBD simulation. As the evolution unfolds, we monitor the behavior of the entanglement entropy between fast and slow degrees of freedom. Once it falls below a pre-established threshold $\eta_S$, we perform the truncation described by the direct decomposition and group the pieces obtained in it into the tensors $M_L, B_\ell$ and $N_R$, as can be seen in Fig.~\ref{fig:schemeHeuristic}. Lastly, we use these tensors as a starting guess for the heuristic optimization described in the paragraph above, which modifies them slightly to correct for the small deviation in the reduced density matrices. Once the optimization is stopped, we substitute the pieces for all the tensors $C$ in our chain, normalize again the state and continue the evolution.
After the tensors $C$ are substituted by their lower-rank counterparts for the first time, 
the structure of the uMPS describing the purification contains a sequence of tensors $B_{\ell}$ that carry the physical indices, separated by two \emph{purification} tensors $N_R$ and $M_L$, themselves connected by a diagonal of inverse Schmidt values.
Notice that the subsequent time evolution is applied only to the physical indices of each $B_{\ell}$ tensor, but not to the purification ones.
Nevertheless, these tensors mediate the entanglement between physical sites, which in general will continue building up in time. 
We thus need to continue monitoring the  
fast-slow entanglement entropy of the $B_{\ell}$ tensors in order to apply further truncations as required. 

Two extra aspects of the algorithm deserve a comment. The first one is that 
with each truncation/mixing step, the number of purification sites between a pair of blocks increases by two. 
In principle, since there is a large freedom in the representation of the purification indices, 
it would be possible to look for a more efficient description of this set of tensors (e.g. minimally entangled, in the spirit of~\cite{Hauschild2018puri}).
In this work, for simplicity, 
we group all the purification legs together and work with a unit cell that has two tensors, one with the physical dimension and another one with the purification dimension. The dimension of the purification index would thus in principle grow exponentially with the number of truncations.
In order to efficiently deal with it, we
perform a SVD of the purification tensor, splitting virtual vs purification legs, and
implement a hard cutoff on this dimension
(with a value of a $d_{\textrm{purification}} = 1000$, for our simulations). 
Then we continue the evolution while keeping the purification dimension fixed. 
The second aspect that deserves a mention refers to the performance of the heuristic truncation without using the input from the entanglement decomposition. We observe that whenever we do not make use of it, the heuristic optimization results (i.e. the distance between the correct marginals and the density matrices of the varitional mixed state) are not good, and thus, we incur in errors that accumulate during the simulation. 

\section{Measure of long-range entanglement}

In the main text we use the logarithmic negativity \cite{Vidal2002, Plenio} to quantify the entanglement between the left part of the system $L$ and the right part of the system $R$ that is mediated by the long-range degrees of freedom. In this section, for the sake of completeness of our discussion, we briefly introduce its definition and mention our motivation for using it to characterize the long-range entanglement.

Given a bipartite mixed state $\rho_{LR}$, the logarithmic negativity is defined as

\beq
E_N (\rho_{LR}) = \log_2 \lVert \rho_{LR}^{\Gamma_L} \rVert_1, 
\eeq where $\lVert A \rVert_1 = \textrm{Tr} \sqrt{A^\dagger A}$ is the trace norm of a matrix, i.e. the sum of its singular values and $\rho_{LR}^{\Gamma_L}$ denotes the partial transposition with respect to the subsystem $L$. If we write $\rho_{LR}$ in the tensor product basis

\beq
\rho_{LR} = \sum_{ijkl} \rho_{ik,jl} \left( \ket{i}\bra{j} \right)_L \otimes \left( \ket{k}\bra{l} \right)_R,
\eeq then the partial transpose reads

\beq
\rho_{LR}^{\Gamma_L} = \sum_{ijkl} \rho_{ik,jl} \left( \ket{i}\bra{j} \right)_L \otimes \left( \ket{l}\bra{k} \right)_R.
\eeq A non-zero logarithmic negativity indicates that the state under consideration is entangled, and it can be shown that it is a proper entanglement monotone \cite{Plenio}. In the main text we use it to characterize the entanglement mediated by the fast degrees of freedom, as it is useful when the state under consideration is mixed. This is the case in our study, as at finite times the fast degrees of freedom have not decoupled completely from the rest of excitations present in the system and they still carry some residual entanglement with the former.

\section{Dependence on the block size}

\begin{figure}[t]
	\centering
	\includegraphics[width=.9\columnwidth]{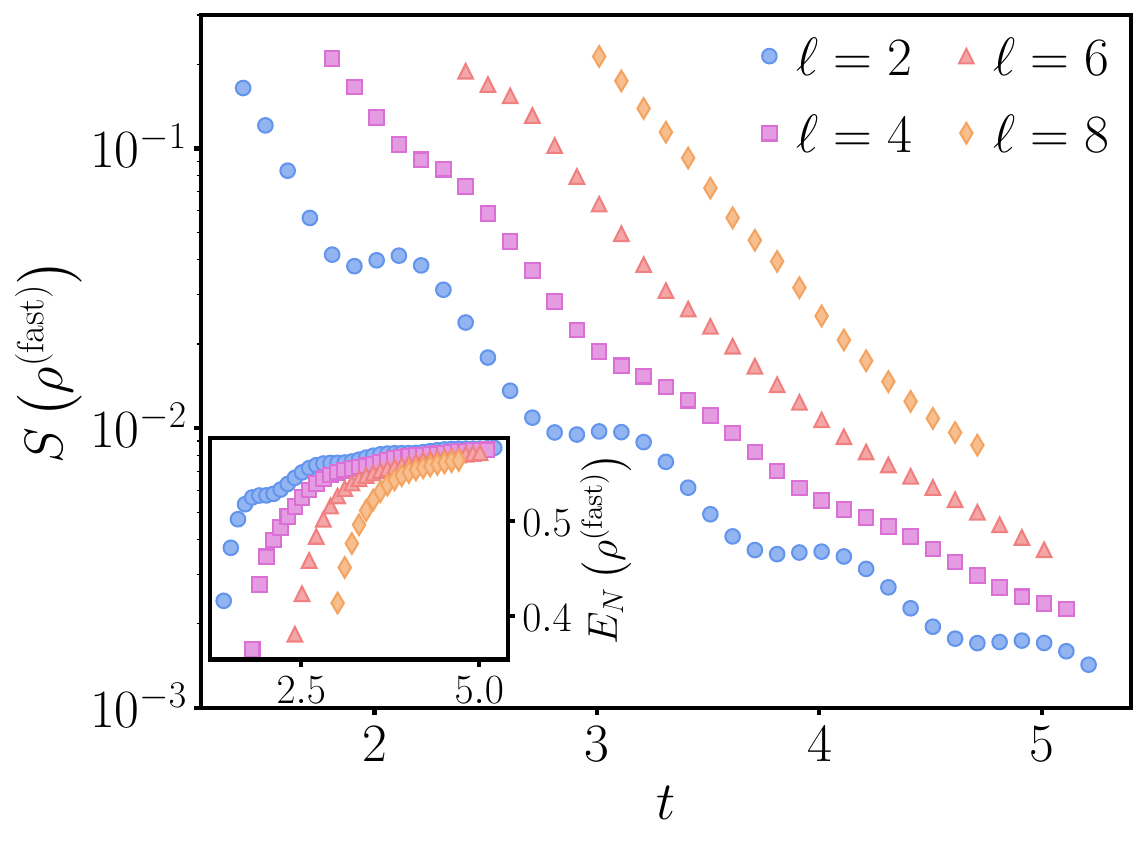}
	\caption{Decay of the entropies (main plot) and behavior of the logarithmic negativity of the fast degrees of freedom (inset) as a function of time for different sizes of the subsystem $S$, $\ell$. The parameters for the quench are $g = 2, J_2 = 0$, the same phase diagram point that we use in the main text.
 }
 \label{fig:figBlockSize}
 \end{figure}

In this section, we show how the results presented in Fig.~\ref{fig:figEntropy} of the main text depend on the size $\ell$ of the subsystem $S$. As can be seen from Fig.~\ref{fig:figBlockSize}, the larger the size of $S$, the longer it takes for the fast degrees of freedom to disentangle. Increasing the size of the subsystem $S$ linearly leads to an apparently linear increase in the timescale in which excitations disentangle from $S$. Still, the decay of the entropy seems to be exponentially fast with time. In addition, if we wait for long enough, the long-range entanglement found by our algorithm seems to saturate to the same value for all $\ell$. This seems to indicate that the same excitations that have decoupled from a subsystem $S$ of size $\ell = 2$ propagate and eventually decouple from larger subsystems, provided we wait long enough.

\section{Free fermion calculations}

In the integrable case ($J_2=0$), the model can be mapped to an exactly solvable free fermionic system. We can use this solution to perform the same calculations as with the TN algorithm, and thus
corroborate and gain a deeper understanding of the observed numerical results.

\subsection{Disentangling the long-range components}

Fig.~\ref{fig:figEntropy}(a) of the main text shows that, for quenches from the ground state at $g\to \infty$, the residual entropy between fast and slow degrees of freedom decays slower for smaller values of $g$, even though for all the cases studied the decay is compatible with an exponentially fast decrease in time.
At the same time, the saturation value of the logarithmic negativity carried by the fast degrees of freedom seems to grow as we scan the final value of the transverse field from $g = 2$ to $g = 0.5$. In fact, at $g = 1$, the logarithmic negativity seems to saturate to $1$, the maximal possible value for a bipartite state of two qubits~\cite{Vidal2002}.

\begin{figure*}[t]
	\centering
	\includegraphics[width=\linewidth]{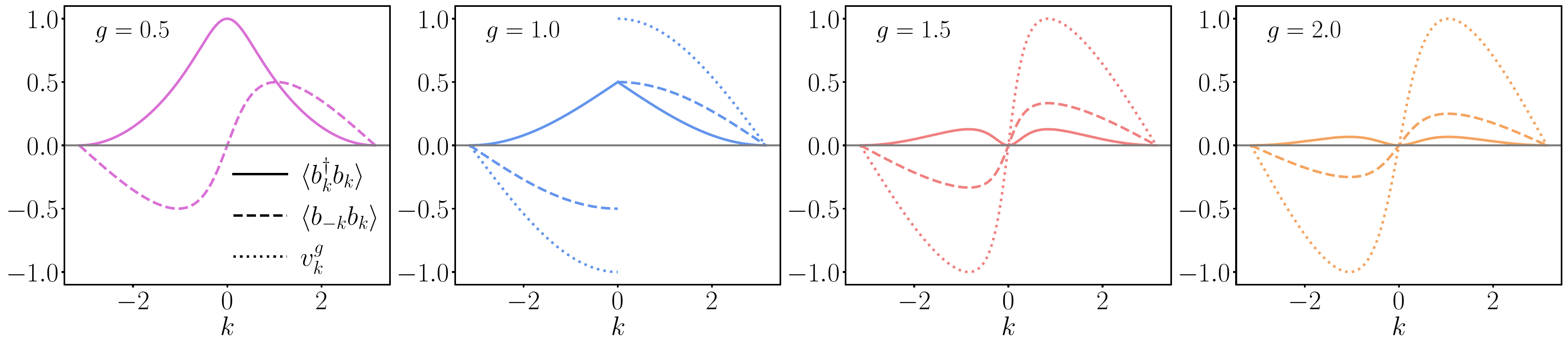}
	\caption{Occupation of the eigenmodes of the quench Hamiltonian (solid lines) and their group velocity (dotted lines) as a function of their momentum in the integrable quenches shown in Fig.~\ref{fig:figEntropy}(a) of the main text. Dashed lines show the initial coherence between opposite momenta. In the leftmost plot ($g = 0.5$), the coherence between opposite momenta and the group velocity are on top of each other.}	\label{fig:fig_supp_mat_1}
\end{figure*}

We can relate this behavior to the 
 distribution of quasi-particles in the initial state, illustrated in figure~\ref{fig:fig_supp_mat_1} for the studied quenches.
As we scan from $g = 2$ to $g = 0.5$ in the phase diagram, the occupation number of the freely propagating modes of the quench Hamiltonian grows. Something similar happens to the coherences between opposite momenta. This is reminiscent of the behavior of the logarithmic negativities in the TN computations and of the fact that as we quench to smaller values of $g$, the energy of the initial state according to the final Hamiltonian gets closer to the middle of the spectrum. Furthermore, by analysing the group velocities one can get some intuition of the rates of decay of the residual entropy. For a quench inside the paramagnetic phase, the maximum of the occupation coincides with the maximal group velocity. In contrast, when we quench to the ferromagnetic phase, the maximally occupied mode has vanishing group velocity. This could be an explanation of the slower decay rate observed when quenching to the ferromagnetic phase.

\subsection{Eliminating the long-range coherences}

\begin{figure*}[t]
	\centering
	\includegraphics[width=\linewidth]{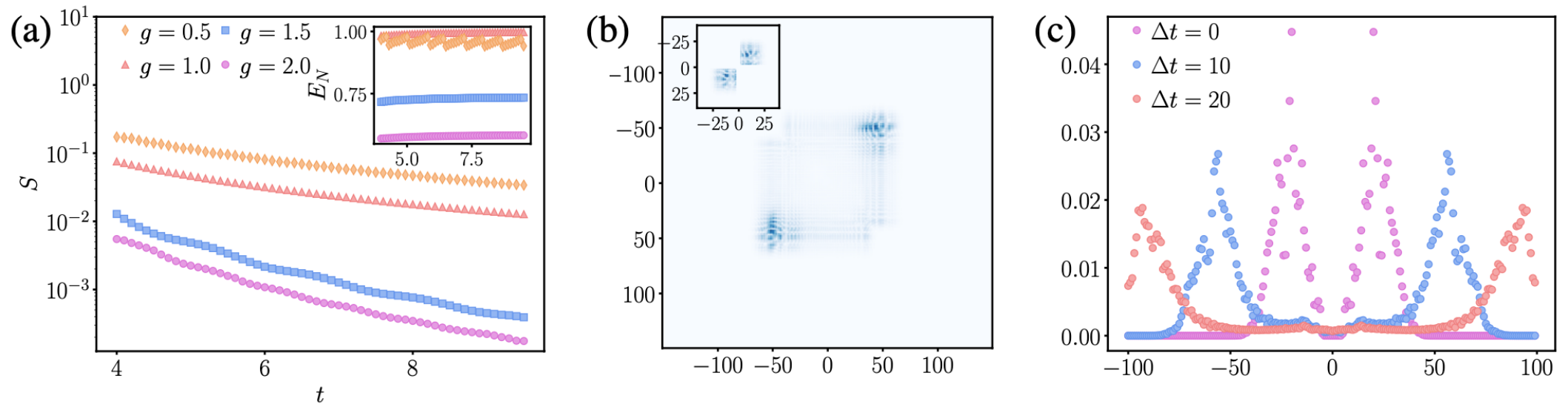}
	\caption{We reproduce the TN calculations for an integrable quench using the free fermionic formalism. For a system of size $N=800$ we consider a block $S$ of 4 sites and identify a pair of modes, respectively in $L$ and $R$, that disentangle from $S$, as explained in the text. The residual fast-slow entropy (a) shows the same behavior as in the TN calculation, as does the logarithmic negativity between the $L$ and $R$ fast components (inset). (b) shows a heatmap of the contribution to the spatial correlation matrix $\langle a^\dagger_i a_j \rangle$ of the coherences of this long-range pair when it is identified at $t=5$ (inset) and after an additional evolution for $\Delta t=20$. Lastly, (c) shows the sum of these contribution as a function of the distance between the insertions of the operators in the correlation matrix.   }
    \label{fig:figFF}
\end{figure*}

The free fermionic chain can also provide insight on the transformation performed by the TN algorithm.
A Jordan-Wigner transformation maps the spin chain to a system of $N$ fermionic modes,  created by operators  $a^{\dagger}_i$, $i=1,\ldots N$, that fulfill standard commutation relations
 $\left\{a^\dagger_i, a_j\right\} =\delta_{i,j}$, with $\delta$ the Kronecker delta. 
Because the resulting Hamiltonian involves only quadratic terms in $a_i$, $a^{\dagger}_j$,
 the fermionic system is free, and ground states and their evolution are described by Gaussian states.
 By virtue of Wick's theorem, these are completely characterised by two-point correlation functions, which can be 
organized in the $2N\times 2N$ correlation matrix~\cite{Surace2022} 
 $\Gamma_{i,j}=\langle \alphaVec_i \alphaVec^{\dagger}_j\rangle$,
 where the  operator-valued vector $$\alphaVec^{\dagger}=\left(a_1,a_2,\dots,a_N,a_1^{\dagger},a_2^{\dagger},\dots,a_{N}^{\dagger}  \right)$$ 
holds all creation and annihilation operators of the chain. Other sets of fermions $\cVec=U\alphaVec$ can be defined through symplectic transformations, which conserve the anticommutation relations.

We can now divide the system in three parts, namely a central region $S$ containing few fermions and playing the role of the local block in the above description,
and two regions $L$ and $R$, containing respectively all the fermions to the left or right of $S$. 
 Because the state is Gaussian, all correlations of operators supported in one connected region are determined by the restriction of the correlation matrix to the modes in that region.
 The symplectic transformation representing a unitary that acts on the left (right) virtual leg of the MPS would in this setting act only on the modes defined in $L$ ($R$).  
 
The free-fermionic equivalent of the TN algorithm should try to single out, after some evolution time, at least one fermionic mode localized in $L$ and entangled only with modes in $R$.
Namely we want to identify modes $l = \sum_j U^L_{1,j}{\alpha}^L_j$  and $r = \sum_j U^R_{1,j}\alpha^R_j$ such that $\langle \lambdaVec \cdot {\alphaVec^S}^{\dagger}\rangle=\langle \rhoVec \cdot {{\alphaVec^S}^{\dagger}} \rangle=0$, where $\rhoVec^{\dagger}=(r,r^\dagger)$, $\lambdaVec^{\dagger} =(l,l^\dagger)$ and $\alphaVec^{S(L,R)}$ is the restriction of $\alphaVec$ to the modes in $S$ ($L$, $R$).  To this end, after discarding modes that are pure in either $L$ or $R$, we optimize numerically the linear combinations above that produce the lowest overlap with modes in $S$. Once these two modes are isolated, we can compute their joint reduced density matrix, study its entanglement properties and compare them to those found in the TN case. This reduced density matrix would be analogous to $\rho_{L_f R_f}^{\mathrm{(fast)}}$. In the context of free fermions, converting the long-range entanglement to mixture is achieved by setting the coherences of the modes, that is, the matrix elements $\langle \rhoVec \cdot {{\lambdaVec}^{\dagger}} \rangle$ to zero. As in the TN case, after the truncation we can continue the evolution and check the effect in the local observables.

In order to compare to the TN results, we have applied this procedure to different quenches for a system of $N=800$ fermionic modes, in which we singled out a central region of four sites. Fig. \ref{fig:figFF}(a) shows the residual entanglement between the $l,\,r$ pair and the modes in $S$ (main panel) for several values of the quench parameter, together with the negativity between the two modes $l$ and $r$ (inset), to be directly compared to the results in Fig.~\ref{fig:figEntropy} of the main text. Both quantities show good agreement with the TN results and confirm that in all scenarios, as time passes, the identification of $l,\,r$ becomes more and more accurate, and that both modes keep being robustly entangled with each other.  Notice that the agreement is not only qualitative, but the magnitude of the entanglement quantities is very similar to those found with the TN algorithm, which supports our interpretation of the TN transformation as correctly identifying the quasiparticles that carry away long-range entanglement.

For the quench $g = 2$, after time $t = 5$, we track the effect of the long-range coherences identified at that time in the observables. Fig.~\ref{fig:figFF}(b) shows (in absolute value) the contribution of the coherences of this pair to the ($a^{\dagger} a$ part of the)  correlation matrix on the spatial basis (inset) as well as how this has evolved after a time $\Delta t=20$ has elapsed. The empty region around $(0,0)$ corresponds to our block $S$. As can be seen from the figure, the coherences only contribute to matrix elements very far away from the diagonal, which correspond to very non-local correlation functions. In contrast, the local expectation values, which are represented in this matrix by the diagonals close to the main one, are practically zero. This fact indicates that the evolution of these coherences does not have an effect on local expectation values, and thus they can be safely discarded, and the state of these modes approximated by a mixture.

As a way to visualize how the contributions of the $l-r$ coherences drift to more and more non-local observables, we show in panel (c), for several instants of time, the sum of the $n$-diagonal of the coherence matrix in Fig.~\ref{fig:figFF}(b) as a function of $n$. This corresponds to the contribution of the $l-r$ coherences to the quantity $\sum_n |\langle a^\dagger_{n+x} a_n \rangle|$, and shows how, as the evolution unfolds, the coherences propagate towards more and more non-local expectation values, never affecting the local ones, as can be seen from the location of the peaks in the figure and the almost vanishing values around the origin.
These findings agree with the ones found by TN, and show that also in the Gaussian formalism it is possible to decohere some of the long-range entanglement generated during a quench while preserving the evolution of the local expectation values.

\end{document}